\documentclass[journal=jctcce,manuscript=article]{achemso}

\usepackage{amsmath}
\usepackage{graphicx}
\usepackage{epstopdf}
\usepackage{xcolor}
\mciteErrorOnUnknownfalse

\author{Andrew M. Sand}
\affiliation{Department of Chemistry, Butler University, Indianapolis, IN 46208}
\email{amsand@butler.edu}
\author{Justin T. Malme}
\author{Erik P. Hoy}
\affiliation{Department of Chemistry and Biochemistry, Rowan University, Glassboro, NJ 08028}
\email{hoy@rowan.edu}

\title{A multiconfigurational pair-density functional theory approach to molecular junctions}

\begin{document}

\begin{tocentry}
\includegraphics[scale=0.5]{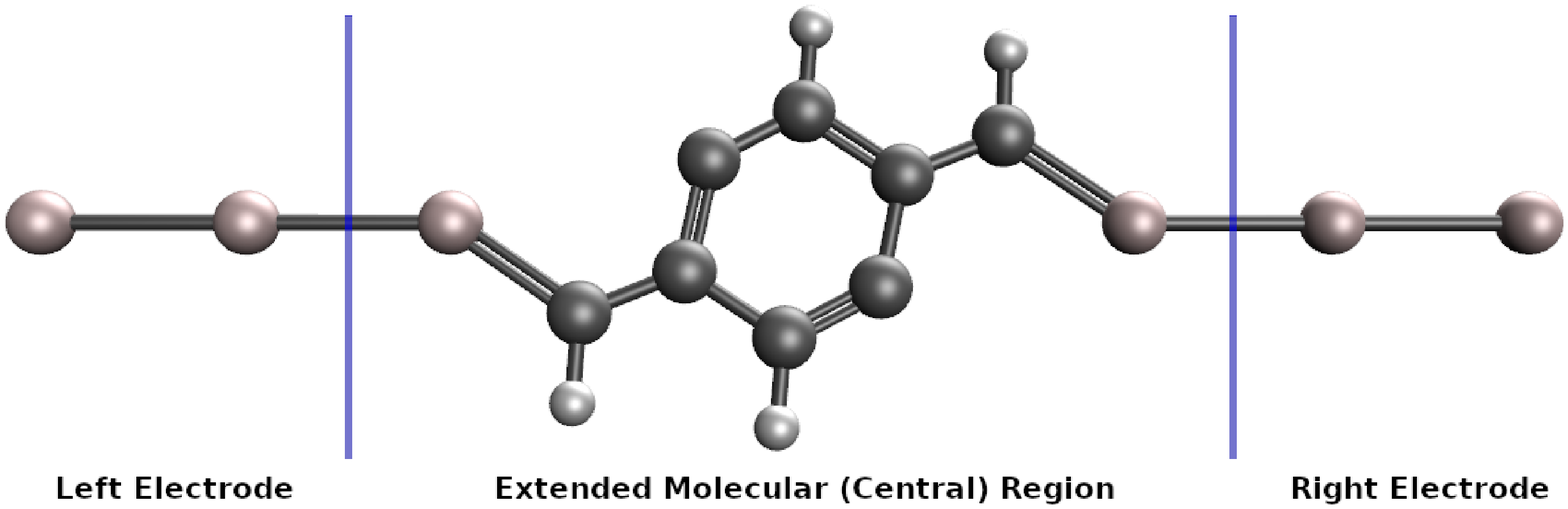}

\end{tocentry}

\begin{abstract}
Due to their small size and unique properties, single-molecule electronics have long seen research interest from experimentalists and theoreticians alike.  
From a theoretical standpoint, modeling these systems using electronic structure theory can be difficult due to the importance of electron correlation in the determination of molecular properties, and this electron correlation can be computationally expensive to consider, particularly multiconfigurational correlation energy. 
In this work, we develop a new approach for the study of single-molecule electronic systems, denoted NEGF-MCPDFT, which combines multiconfiguration pair-density functional theory (MC-PDFT) with the non-equilibrium Green's function formalism (NEGF).  
The use of MC-PDFT with NEGF allows for the efficient inclusion of both static and dynamic electron correlation in the description of the junction's electronic structure.  
CASSCF wave functions are used as references in the MC-PDFT calculation, and like with any active space method, effort must be made to determine the proper orbital character to include in the active space.
We perform conductance and transmission calculations on a series of alkanes (predominantly single-configurational character) and benzyne (multiconfigurational character), exploring the role that active space selection has on the computed results.
For the alkane junctions explored (where dynamic electron correlation dominates), the MCPDFT-NEGF results agree well with DFT-NEGF results.
For the benzyne junction (which has significant static correlation), we see clear differences in the MCPDFT-NEGF and DFT-NEGF results, and evidence that NEGF-MCPDFT is capturing additional electron correlation effects beyond those provided by the PBE functional.
\end{abstract}

\section{Introduction}

Single-molecule electronics have long captivated researchers due to their unique transport properties and small length-scales, making them a promising alternative to and potentially a replacement for conventional silicon-based semiconductor electronics.~\cite{Nitzan2001,Nitzan2003,Tao2006,Cui2017,Tsutsui2012,VentReview2013,Xiang2016,Evers2020}
Significant effort has been invested in developing theoretical models capable of accurately describing  charge transport in these nanoscale systems.~\cite{Thoss2018a}
Despite notable progress, this has proven to be a difficult challenge. 
This difficulty can be attributed to the large range of physical and chemical effects that need to be modeled including molecule-metal interactions, thermal effects, geometric variations, bulk-phase electrode properties, and the electronic structure of the electronic device itself.
Considering all possible interactions and properties for these systems rigorously is not computationally feasible, and thus some properties have received less attention over the years. 
One such property is molecular electronic correlation, which consists of dynamic and multireference (or static) correlation. 
Obtaining a reasonable and balanced description of both types of electron correlation is vital to achieving an accurate description of certain electronic properties such as charge transport. 
While electron correlation effects are well-studied in molecular quantum mechanics, they have received less attention in molecular electronics studies. 
However, treating electronic correlation effects, particularly multireference correlation, in charge transport has become both increasingly important and tractable due to improving computational resources. 
Recent studies have suggested that multi-reference correlation may play a role in determining novel charge transport phenomena such as reversed electrical conductance decay with increasing molecular length, sometimes referred to as anti-ohmic conductance.~\cite{Gil-Guerrero2019,Gil-Guerrero2019,Ramos-Berdullas2021}

Most charge transport studies in molecular electronics employ methods based on non-equilibrium Green's function theory (NEGF) which, in principle, provides an exact description of charge transport.~\cite{NEGF,Nitzan2001,Ratner2007,Xue2002,Stokbro2003,Papior2017,SMEAGOL,NEGF_1DEl,NEGF_TurboMole,NEGF_ADF,Kiguchi2016} 
All practical NEGF methods, however, employ approximate electronic structure methods to construct the 
Green's functions for the NEGF transport equations. 
The majority of NEGF studies employ a single-reference electronic structure method such as density 
functional theory (NEGF-DFT). 
As a result, these NEGF approaches are typically very limited in their ability to treat multireference 
effects. Only a very limited number of studies have integrated a multireference approach such as configuration interaction approaches within the NEGF transport formalism.~\cite{Albrecht2006} 
To the authors' knowledge, none are yet in widespread use even though interest in other multireference transport methods such as multilayer multiconfiguration time-dependent Hartree method, time-dependent reduced density matrix theories, and time-dependent CI theories is growing rapidly.~\cite{Greer2020,ArchubiCD2007Mmfm,RamakrishnanRaghunathan2020Csaq,Sajjan2018,Wang2015,Thoss2018a,Dzhioev2014,Wang2003} 
This is likely due, at least in part, to the computational cost of using a CI wavefunction to construct the Green's function directly. 
This cost can be mitigated if an active space approach is employed, such as the complete active space self-consistent field~(CASSCF) method.
Typically, though, a post-SCF step must be used to recover missing dynamic correlation energy outside of the active space, such as second- order perturbation theory (i.e. CASPT2).~\cite{CASPT2} 
As most molecular junctions use metal electrodes, a perturbation theory correction to the wave function runs a high risk of deconvergence due to the metallic bonding.~\cite{Gruneis2010}
What is needed is a multiconfigurational approach that includes a cost-effective, non-perturbative dynamic correlation correction suitable for treating organo-metallic molecular junctions. 
In this paper, we develop a formalism that addresses this issue by integrating multiconfiguration pair-density functional theory~(MC-PDFT) into the NEGF formalism.


Multiconfiguration pair-density functional theory (MC-PDFT) is an approach to the handling of multiconfigurational molecular systems which combines ideas from multiconfigurational wave function theories with density functional theory~\cite{MCPDFT,ghosh_2017}.  
In an MC-PDFT calculation, a special type of density functional, called an on-top functional, is used to calculate electronic energies using the one- and two-electron reduced density matrices produced from a reference multiconfigurational wave function calculation.  
The on-top functional determines energy contributions using densities and on-top densities (which is related to the probability of finding two electrons in the same region of space) which are determined from the one- and two-electron reduced density matrices.  
Most typically, a CASSCF~\cite{CASSCF} wave function calculation is used, but the MC-PDFT procedure can be used in conjunction with any reference calculation that yields appropriate reduced density matrices.~\cite{EDP_var_pdft,Soumen_GAS,Presti_RAS,mcpdft_zhou}  
MC-PDFT has advantages over other post-SCF multiconfigurational methods including its low cost.~\cite{Sand_algorithm} 
MC-PDFT has shown much success in the treatment of small- to medium-sized chemical systems, determining 
reaction energetics with an accuracy similar to CASPT2.~\cite{rhodopsin,ghosh_2017,MCPDFT_GRAD,Chatt,sharma_mno4}

In this paper, we seek to apply the cost savings and accurate determination of correlation energy afforded 
by MC-PDFT to the area of molecular junctions.  
As a first approach, we combine the MC-PDFT and NEGF formalisms into a new method (NEGF-MCPDFT). To 
evaluate the effectiveness of the NEGF-MCPDFT approach, we examined two test molecular junctions using 
both MC-PDFT and a comparable standard DFT method. 
These junctions are a series of dithiol alkane chains and a benzyne biradical, both attached to linear aluminum electrodes. 
Dithiol alkane chains are well-studied both theoretically and experimentally while the benzyne biradical is a well-known multireference system making them desirable test systems.
The alkane junctions represent a low multireference case and are known to exhibit an exponential decrease in conductance with an increase in the molecular chain length.~\cite{Reed2003,Tao_a2006,Reed2004,Pauly2016,cond_alkane2,Shen2012,cond_alkane1,Nijhuis2015,Fagas2007,
Venkataraman2006a,Kim2013}
The alkane junctions were selected to ensure that MC-PDFT is able to reproduce known molecular trends that can be captured with standard DFT functionals.  
In contrast, the benzyne biradical represents a molecular system exhibiting high multireference correlation.~\cite{Crawford2001,Schutski2014,McManus2015,Ray2016}
This system was chosen to serve as a test to evaluate if the inclusion of multi-reference character from an MC-PDFT calculation makes a significant difference in both the predictions of transmission and currents in the junction. 
In the more multireference systems, we expect to find the greatest deviation between the standard DFT functionals and MC-PDFT. 

\section{Theory}
In this section, we outline the development of the non-equilibrium Green's function pair density functional theory (NEGF-MCPDFT) method.
We first review the NEGF and MC-PDFT theories before illustrating how these theories are integrated to form the NEGF-MCPDFT method.

\subsection{Non-equilibrium Green's Function} \label{nonequil}

\begin{figure}
\includegraphics[scale=0.5]{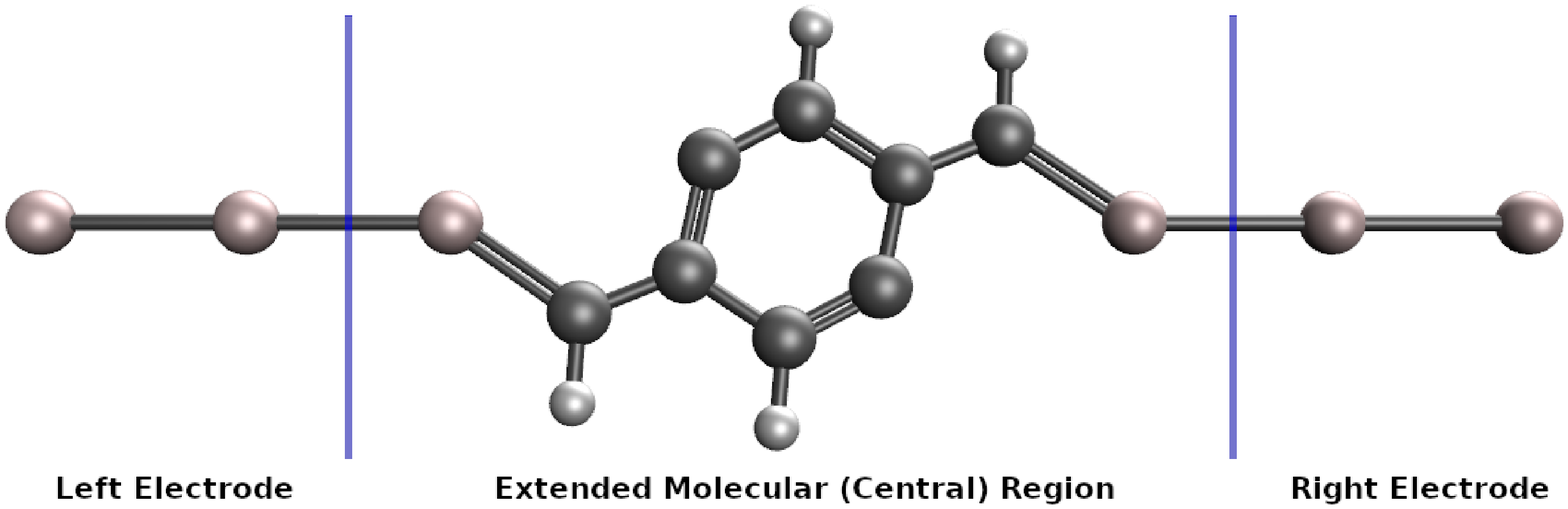}
\caption{The partitioning scheme used for the benzyne radical junction.} 
\label{ben_jxn_part}
\end{figure}

In most molecular junction calculations employing the non-equilibrium Green’s function formalism, the standard approach is to partition the transport problem into extended molecular, left electrode, and right electrode regions as shown in Figure~\ref{ben_jxn_part}.~\cite{DFT1,DFT2,DFT3,GW2012} 
If the transport region is in a steady-state and the electrode voltage difference is small, then the Green's functions for the molecular and electrode regions can be obtained from a ground state electronic structure method. 
The retarded Green's function for the extended molecular region is defined as
\begin{equation}
G_\mathrm{EM} = [ES_\mathrm{EM} - H_\mathrm{EM} - \Sigma_\mathrm{L}(E) - \Sigma_\mathrm{R}(E)]^{-1}
\label{eq:RGF}
\end{equation}
where $E$ is the energy, $S_\mathrm{EM}$ and $H_\mathrm{EM}$ are the overlap matrix and Hamiltonian matrix in the extended molecular (central) region, respectively, and $\Sigma_\mathrm{L}(E)$ and $\Sigma_\mathrm{R}(E)$ are the self energies that describe the interactions between each electrode and the extended molecular region.

For low-bias junctions, the current values for the molecular junction are then typically obtained using the Landauer formula for ballistic conductance,
\begin{equation}
I =  \frac{2e^2}{h} \int T(E)[f_L(E)-f_R(E)]dE.
\end{equation}
where T is the transmission function,
\begin{equation}
T(E) = Tr(\Gamma_{\mathrm{R}}\:G^{R}_{\mathrm{EM}}\Gamma_{\mathrm{L}}\:G^{A}_{\mathrm{EM}}),
\end{equation}
and $f_\mathrm{L}(E)$ and $f_\mathrm{R}(E)$ are the Fermi functions for the left and right electrodes and $\:G^R_{\mathrm{EM}}$ and $
\:G^A_{\mathrm{EM}}$ are the retarded and advanced Green's functions for the extended molecular region. 
For this work, we employed a Landauer NEGF formalism, although the formalism discussed below could potentially be applied to a self-consistent NEGF formalism.~\cite{Landauer1957}

\subsection{Multiconfiguration Pair-density functional theory}

Multiconfiguration pair-density functional theory is a method that combines the advantages of multiconfiguration wave function theory with density functional theory.
In a typical MC-PDFT calculation, a state-specific or state-averaged CASSCF wave function is used as a reference, and the MC-PDFT energy of each state is determined by the following equation:
\begin{equation}
E_\mathrm{MC-PDFT} = V_{nn} + \sum_{pq} h_{pq} D_{pq} + \frac{1}{2}\sum_{pqst} g_{pqst} D_{pq} D_{st} + E_\mathrm{ot}[\rho ({\bf{r}}),\Pi ({\bf{r}}),\rho ' ({\bf{r}}),\Pi '({\bf{r}})]
\label{eq:pdft_energy}
\end{equation}
where $p,q,s,t$ indicate general orbital indices, $\bf{r}$ is a three-coordinate spatial vector, $V_{nn}$ is the nuclear repulsion energy, $h_{pq}$ and $g_{pqst}$ are the one- and two-electron integrals, $D_{pq}$ is the one-body reduced density matrix of the state of interest, and $E_\mathrm{ot}$ is the on-top density functional which includes the electron density $\rho ({\bf{r}})$ and on-top pair density $\Pi ({\bf{r}})$ (and their spatial derivatives) of the particular electronic state as arguments:
\begin{equation}
\rho({\bf{r}}) = \sum_{pq} \phi_p({\bf{r}}) \phi_q({\bf{r}}) D_{pq}
\end{equation}
\begin{equation}
\Pi({\bf{r}}) = \sum_{pqst} \phi_p({\bf{r}}) \phi_q({\bf{r}})\phi_s({\bf{r}}) \phi_t({\bf{r}}) d_{pqst}
\end{equation}
where $\phi$ are electron orbitals.
The on-top density functionals used in this work are formed through the translation of the input densities to existing generalized gradient approximation Kohn-Sham (KS) density functionals which depend on the spin-up and spin-down electron densities $\rho_{\alpha}$ and $\rho_{\beta}$ as well as their gradients $\rho'_{\alpha}$ and $\rho'_{\beta}$.  The translation scheme is given by
\begin{equation} 
\tilde{\rho}_{\alpha}({\bf{r}}) = 
\begin{cases}
\frac{\rho({\bf{r}})}{2}(1+\zeta_t({\bf{r}})) & R({\bf{r}}) \leq 1 \\
 \frac{\rho({\bf{r}})}{2} & R({\bf{r}}) > 1
 \end{cases}
 \label{eq:t_start}
\end{equation}
\begin{equation} 
\tilde{\rho}_{\beta}({\bf{r}}) = 
\begin{cases}
\frac{\rho({\bf{r}})}{2}(1-\zeta_t({\bf{r}})) & R({\bf{r}}) \leq 1 \\
 \frac{\rho({\bf{r}})}{2} & R({\bf{r}}) > 1
 \end{cases}
\end{equation}
\begin{equation} 
\tilde{\rho}_{\alpha}'({\bf{r}}) = 
\begin{cases}
\frac{\rho'({\bf{r}})}{2}(1+\zeta_t({\bf{r}})) & R({\bf{r}}) \leq 1 \\
 \frac{\rho'({\bf{r}})}{2} & R({\bf{r}}) > 1
 \end{cases}
\end{equation}
\begin{equation} 
\tilde{\rho}_{\beta}'({\bf{r}}) = 
\begin{cases}
\frac{\rho'({\bf{r}})}{2}(1-\zeta_t({\bf{r}})) & R({\bf{r}}) \leq 1 \\
 \frac{\rho'({\bf{r}})}{2} & R({\bf{r}}) > 1
 \end{cases}
\end{equation}
where the intermediates are defined as
\begin{equation}
\zeta_t({\bf{r}}) = \sqrt{1-R({\bf{r}})}
\end{equation}
\begin{equation}
R({\bf{r}}) = \frac{\Pi({\bf{r}})}{[\rho({\bf{r}})/2]^2}
\end{equation}
\begin{equation}
\rho({\bf{r}}) = \rho_{\alpha}({\bf{r}}) + \rho_{\beta}({\bf{r}}) 
\end{equation}
\begin{equation}
\rho'({\bf{r}}) = \rho'_{\alpha}({\bf{r}}) + \rho'_{\beta}({\bf{r}})
\label{eq:t_end}
\end{equation}

\subsection{NEGF-MCPDFT}
In order to integrate multi-reference effects into NEGF methods, we developed a formalism for combining the retarded Green's function with PDFT methods (NEGF-MCPDFT). 
In the NEGF-MCPDFT approach, the Green's function (eq. \ref{eq:RGF}) uses the MC-PDFT energy determined via eq. \ref{eq:pdft_energy} and the orbital overlap matrix from the reference CASSCF calculation.  
The definition of the Hamiltonian requires careful discussion.  
In a typical NEGF-DFT approach, this Hamiltonian matrix takes the form of a Kohn-Sham matrix, which is single-particle in nature.  
The MC-PDFT method, however, inherently requires elements from the two-particle reduced density matrix in order to determine the on-top density required by the on-top density functional.  
However, in our current approach, we neglect the effects of the two-electron reduced density matrix and we define a pseudo-Fock matrix analogous with Kohn-Sham matrix from DFT, determined by taking partial derivatives of the MC-PDFT energy expression with respect to the 1-body reduced density matrix elements:
\begin{equation}
F_{pq} = h_{pq} + \frac{1}{2}\sum_{pqst}g_{pqst}D_{st} + \frac{\partial E_\mathrm{ot}[\rho,\Pi,\rho ', \Pi ']}{\partial D_{pq}}.
\end{equation}
The last term, denoted the one-electron on-top potential, is calculated in terms of the derivatives of the KS density functional in conjunction with the translation scheme described in equations \ref{eq:t_start}-\ref{eq:t_end}.  In an orbital basis, this term is expressed as
\begin{multline}
\frac{\partial E_\mathrm{ot}[\rho,\Pi,\rho ', \Pi ']}{\partial D_{pq}} =
\frac{\partial E_{\mathrm{xc}}[\tilde{\rho}_{\alpha},\tilde{\rho}_{\beta},\tilde{\rho}_{\alpha}',\tilde{\rho}_{\beta}']}{\partial \tilde{\rho}_{\alpha}} \frac{\partial \tilde{\rho}_{\alpha}}{\partial D_{pq}} +
\frac{\partial E_{\mathrm{xc}}[\tilde{\rho}_{\alpha},\tilde{\rho}_{\beta},\tilde{\rho}_{\alpha}',\tilde{\rho}_{\beta}']}{\partial \tilde{\rho}_{\beta}} \frac{\partial \tilde{\rho}_{\beta}}{\partial D_{pq}} \\ +
\frac{\partial E_{\mathrm{xc}}[\tilde{\rho}_{\alpha},\tilde{\rho}_{\beta},\tilde{\rho}_{\alpha}',\tilde{\rho}_{\beta}']}{\partial \tilde{\rho}_{\alpha}'} \frac{\partial \tilde{\rho}_{\alpha}'}{\partial D_{pq}} + 
\frac{\partial E_{\mathrm{xc}}[\tilde{\rho}_{\alpha},\tilde{\rho}_{\beta},\tilde{\rho}_{\alpha}',\tilde{\rho}_{\beta}']}{\partial \tilde{\rho}_{\beta}'} \frac{\partial \tilde{\rho}_{\beta}'}{\partial D_{pq}} \label{eq:pot1.2}
\end{multline}
\begin{align}
 \frac{\partial \tilde{\rho}_{\alpha}({\bf r})}{\partial D_{pq}} &= 
 \begin{cases}
 \bigg(\frac{1}{2}\Big[1+\zeta_t({\bf r})\Big]+\frac{2\Pi({\bf r})}{\zeta_t({\bf r})[\rho({\bf r})]^2}\bigg)\Big[\phi_p({\bf r})\phi_q({\bf r})\Big] &  R({\bf{r}}) \leq 1 \\
 \frac{1}{2}\Big[\phi_p({\bf r})\phi_q({\bf r})\Big] & R({\bf{r}}) > 1
 \end{cases}  \\
 \frac{\partial \tilde{\rho}_{\beta}({\bf r})}{\partial D_{pq}} &= 
  \begin{cases}
 \bigg(\frac{1}{2}\Big[1-\zeta_t({\bf r})\Big]-\frac{2\Pi({\bf r})}{\zeta_t({\bf r})[\rho({\bf r})]^2}\bigg)\Big[\phi_p({\bf r})\phi_q({\bf r})\Big] &  R({\bf{r}}) \leq 1 \\
 \frac{1}{2}\Big[\phi_p({\bf r})\phi_q({\bf r})\Big] & R({\bf{r}}) > 1
 \end{cases} \\
 \frac{\partial \tilde{\rho}_{\alpha}'({\bf r})}{\partial D_{pq}} &= 
  \begin{cases}
 \frac{1}{2}\Big[1+\zeta_t({\bf r})\Big]\Big[\phi_p'({\bf r})\phi_q({\bf r}) + \phi_p({\bf r})\phi_q'({\bf r})\Big] + \frac{2\rho'({\bf r})\Pi({\bf r})}{\zeta_t({\bf r})[\rho({\bf r})]^3}\Big[\phi_p({\bf r})\phi_q({\bf r})\Big] & R({\bf{r}}) \leq 1 \\
 \frac{1}{2}\Big[\phi_p'({\bf r})\phi_q({\bf r}) + \phi_p({\bf r})\phi_q'({\bf r})\Big] & R({\bf{r}}) > 1
 \end{cases} \\
 \frac{\partial \tilde{\rho}_{\beta}'({\bf r})}{\partial D_{pq}} &= 
  \begin{cases}
  \frac{1}{2}\Big[1-\zeta_t({\bf r})\Big]\Big[\phi_p'({\bf r})\phi_q({\bf r}) + \phi_p({\bf r})\phi_q'({\bf r})\Big] - \frac{2\rho'({\bf r})\Pi({\bf r})}{\zeta_t({\bf r})[\rho({\bf r})]^3}\Big[\phi_p({\bf r})\phi_q({\bf r})\Big] & R({\bf{r}}) \leq 1 \\
 \frac{1}{2}\Big[\phi_p'({\bf r})\phi_q({\bf r}) + \phi_p({\bf r})\phi_q'({\bf r})\Big] & R({\bf{r}}) > 1
 \end{cases} 
\end{align}

These terms have been previously derived in the context of MC-PDFT analytical gradients~\cite{MCPDFT_GRAD}.
The Kohn-Sham-like matrix is then transformed from the molecular orbital (MO) basis to the atomic orbital (AO) basis
\begin{equation}
{\bf{F}}_\mathrm{AO} = {\bf{SCF}}_\mathrm{MO}{\bf{C}}^T
\label{eq:fock}
\end{equation}
where ${\bf{S}}$ is the overlap matrix and ${\bf{C}}$ is the matrix containing the MO coefficients. 

This approach allows for the 1-particle Green's function to be generated in the same manner as in the NEGF-DFT methodology discussed in Section~\ref{nonequil}. 
The key difference is that the Kohn-Sham Fock matrix is replaced with the approximate PDFT Fock matrix. 
The inclusion of the on-top potential with the CASSCF reference allows for descriptions of both multireference correlation and dynamic correlation to be included in the NEGF formalism without modifying the underlying NEGF approach making it applicable to many existing NEGF algorithms.

\section{Computational Methodology}

The calculations in this work were performed using three software packages: OpenMolcas~\cite{OpenMolcas}, Q-Chem~\cite{Qchem}, and Rowan University Quantum Transport (RUQT)~\cite{RUQT}. Optimized geometries for the molecular regions of the alkane chains were obtained using KS-DFT with the B3LYP functional and the 6-31G* basis set via the Q-Chem software package.~\cite{Hoy2017} The benzyne geometry was optimized using KS-DFT with the PBE functional and the cc-pVDZ basis set in Q-Chem by augmenting the molecule with hydrogen-capped aluminum atoms at each end.  
The MC-PDFT calculations required for the transport calculations were performed using a custom implementation of the MC-PDFT algorithm 
currently available as a branch of the OpenMolcas Github. The tPBE functional was used with all MC-PDFT calculations. 
The standard DFT calculations required for NEGF-DFT transport calculations were performed using the PBE functional within Q-Chem. 
Using the Fock/Overlap matrices generated from the OpenMolcas/Q-Chem calculations, wide-band limit NEGF transport calculations were performed for both methods using the RUQT software package available on Github.~\cite{Hoy2017,RUQT,NEGF,WBL,Kiguchi2016}
All transport calculations utilized the cc-pVDZ basis set, and no spatial symmetry was employed for any of the calculations. 

\section{Results and Discussion}

\subsection{Alkane Conductance with Length}
\begin{figure}
\includegraphics[scale=1]{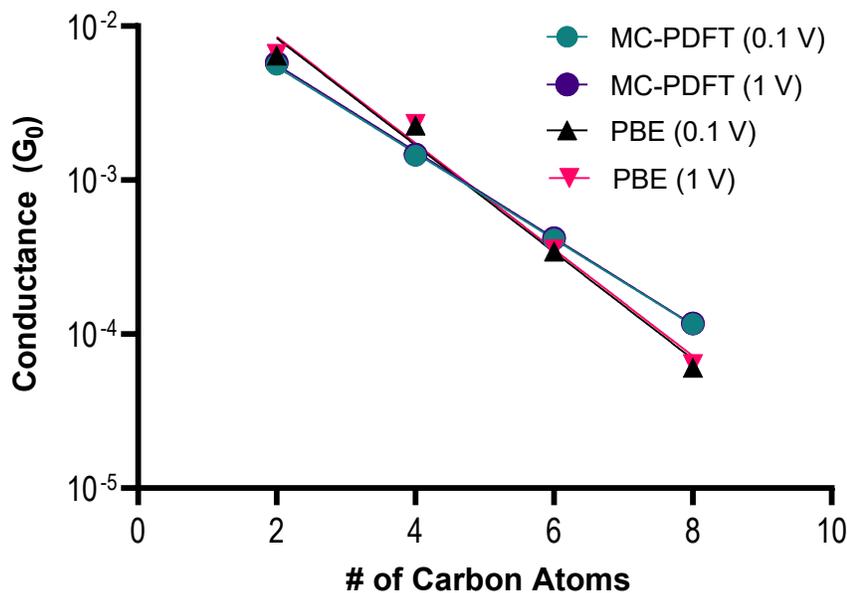}
\caption{The conductance (in quantum of conductance) vs. alkane chain length for both the NEGF-MCPDFT calculations with a (10e, 10o) active space and the NEGF-DFT calculations with the PBE functional. The Exponential fits are shown with solid lines.} 
\label{Figure_cond_alkane_length}   
\end{figure}

It has been well-established through physical experimentation that the conductance in a molecular junction decreases with an increase in molecular length in common molecular resistors like dithiol alkane chains.~\cite{Reed2003,Reed2004,Tao_a2006,Tao_t2006,deboer2008,Shen2012,Nijhuis2015}
To verify that this trend can be reproduced via theoretical NEGF-MCPDFT calculations, we performed a series of calculations on dithiol alkane chains of increasing length.  
While most previous experimental and theoretical studies have used gold electrodes, we employed aluminum electrodes for this test to control computational costs and maintain a single basis set for the entire system without employing an effective core potential.  Although this likely will reduce the agreement with past conductance values and/or trends, we do not expect this change to result in a significant difference in the relationship between the predicted length versus conductance trends of each method.  For the MC-PDFT calculations, we used CASSCF reference wave functions with a (10e, 10o) active space for all alkane chains in the series. The active space orbitals were chosen by considering the SCF orbital energies and selecting the occupied and virtual orbitals nearest to the HOMO-LUMO gap. 

The results of these calculations are shown in Figure~\ref{Figure_cond_alkane_length}. At both 0.1 and 1 
V, we see that the NEGF-MCPDFT and NEGF-DFT methods both predict a decrease in conductance with molecular length. 
 For each method, we have fitted the results to an exponential function, $a*e^{-\beta*x}$, in 
order to compare the trends to experiment. The exponential constant, 
$\beta$, is well-studied experimentally with an accepted value of around 0.92 +/- 0.19 as determined by 
Akkerman and de Boer by averaging the previous experimental results for gold electrodes.~\cite{deboer2008} 
As they noted, however, a number of experimental $\beta$ values have been attained that are significantly outside 
of the margin of error for this value.
When considering our results, we find that the $\beta$ values are both $\sim 0.64$ for NEGF-MCPDFT at 0.1~V and 
1~V, respectively. 
The corresponding values for PBE are both $\sim 0.83$ at 0.1~V and 1.0~V. The PBE values are consistent with the typical values of past DFT calculations using hybrid functionals (0.8-0.95). The NEGF-MCPDFT beta values fall below the bottom of the range for gold electrodes (0.73). 
Comparing the linearity of the curves, the MC-PDFT decay is highly linear $(R^2=0.999)$ while the PBE decay fit is somewhat less linear $(R^2=0.94)$.
Overall, both methods provide reasonably consistent results with the difference between them not particularly large. 

These results show noticeable differences in the predicted decay rate but with the same trend. Compared to 
the results for gold, PBE functional $\beta$ value is closer to the experimental result for gold electrodes 
but both are within the range of previous predictions. Both voltages give very similar conductance values 
indicting that the charge transport is non-resonant and that the Fermi-level and orbitals in the transport 
region are not close in energy. The predicted conductance values of both methods are in close quantitative 
agreement with each other and are within one order of magnitude. As alkane chains do not display 
significant multireference character, we would expect this result. The lack of multi-reference correlation 
means that the largest difference between the two methods is the difference in the effective orbital 
energies. 
The MC-PDFT calculations use CASSCF-optimized orbitals while the DFT calculations use DFT-optimized orbitals. 
In the case of alkane chains, overall we find minor differences in the predicted trends and highly similar 
conductance values for these systems.

\subsection{Benzyne: Active Spaces}

\begin{figure}
\includegraphics[scale=0.15]{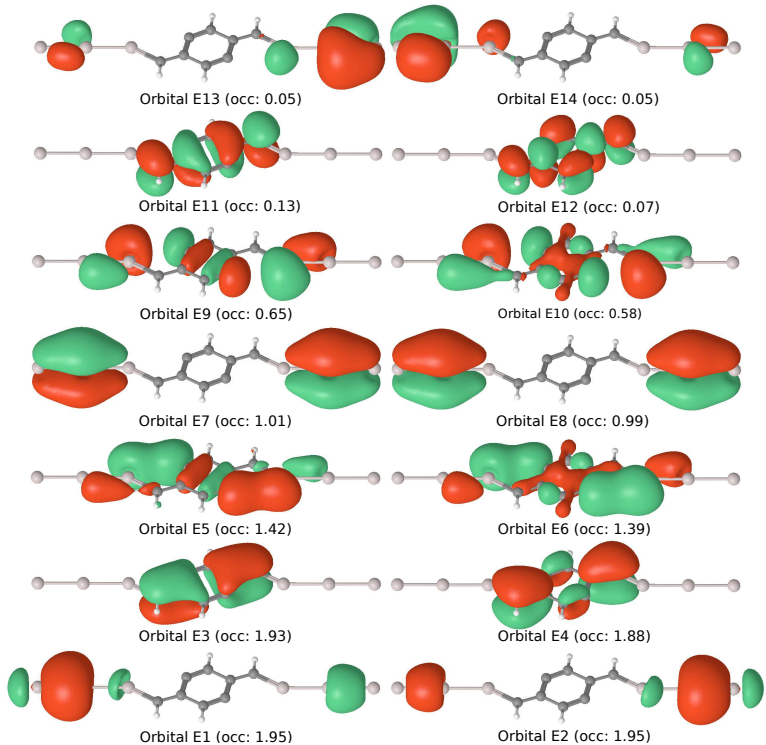}
\caption{Active space orbitals used in the (14e,14o) MC-PDFT energetic orbital set calculations.  Orbital occupation numbers are given in parentheses.} 
\label{Figure_orbitals_1}   
\end{figure}
\begin{figure}
\includegraphics[scale=0.15]{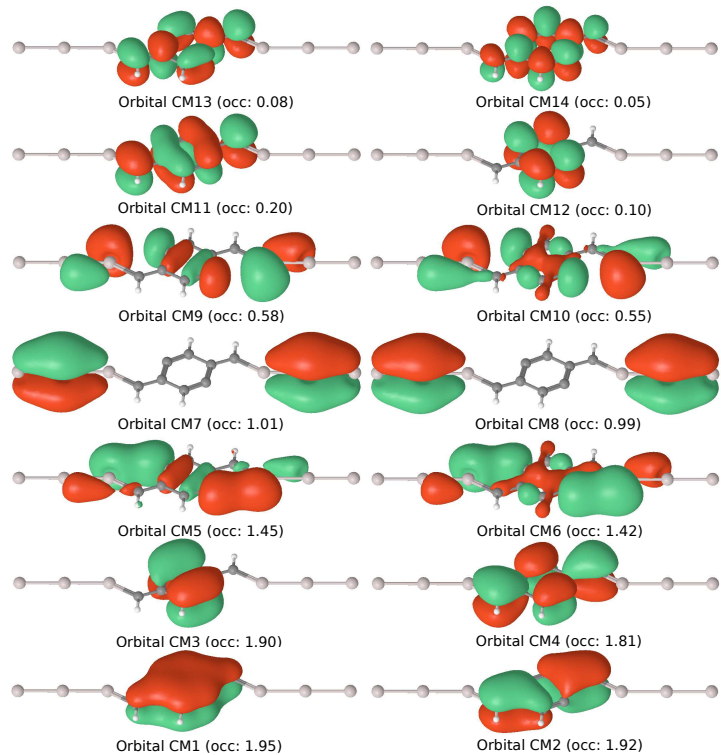}
\caption{Active space orbitals used in the (14e,14o) MC-PDFT central-molecular calculations.  Orbital occupation numbers are given in parentheses.} 
\label{Figure_orbitals_2}   
\end{figure}
To examine the performance of the NEGF-MCPDFT method in a system exhibiting strong multi-reference correlation, we characterized the transmission and conductance of a benzyne biradical molecular junction 
with a carbon linking group. To do this with the MC-PDFT approach, one must first establish a representative active space for this 
system. Throughout this work, all NEGF-MCPDFT calculations make use of an underlying CASSCF reference calculation.  A key step in any CASSCF calculation is the determination of the size and orbital character of the active space.  
A significant number of previous works have focused on schemes for active space selection in order to obtain better reaction energetics or other properties of molecular systems.~\cite{Tischenko2008,Slavicek2010,Veryazov2011,Keller2015,Stein2016,Smith2017,sharma2018,Levine2019,Lischka2020,Cardenas2021}  
Simulations of molecular junctions using a CASSCF approach may require alternative active space selection schemes which are different than the better-studied molecular-based schemes.  
For the benzyne junction, we employed active spaces that ranged in size from (6e,6o) to (14e,14o), and we also explored two 
different active orbital selection schemes.
A sample of the orbitals used in each of the two active space types is shown in Figures~\ref{Figure_orbitals_1} and~\ref{Figure_orbitals_1}.  
We refer to the first set as the ``energetic'' orbital set, which is obtained by seeding the active space CASSCF calculation with the canonical set of Hartree-Fock orbitals, selecting the orbitals with energies closest to the HOMO-LUMO gap. The resultant CASSCF optimized orbitals are then used in the subsequent MC-PDFT calculation step. This orbital set showed a few key features, namely that the HOMO and LUMO orbitals (E6, E7) are always 
centered on the electrodes and display half occupancy as expected. 
For all active space sizes using the energetic orbital set, the HOMO and LUMO appear to be aluminum $\pi$-type orbitals. 
The CASSCF optimization consistently favored including these 
electrode-centered $\pi$-type orbitals over any molecular-centered orbitals when given this initial seeding.  
The inclusion of significant electrode-centered orbital character in the active space is to be expected as partial occupancy (which the active space allows) is required for an adequate description of the linear three-Al atom electrodes. 
This approach resulted in CASSCF active space orbitals 
that tended to exhibit a large amount of electrode character. 
This characteristic represents a potential active space design principle for representing a transport system that is defined by the molecule-electrode interactions or a high degree of Fermi-level and molecular orbital alignment. 

We also employed a second set of CASSCF orbitals, denoted the ``central-molecular'' orbital set.
This active orbital set was designed to include $\pi$-like orbitals (and their correlating anti-bonding orbitals) and the orbitals associated with the diradical from the central molecular region along with two orbitals from the electrodes (which must be included to allow for half-filling in the metal atoms).  These orbitals most strongly contribute to the multiconfigurational character of the system.  The (14e,14o) set was obtained by using the (14e,14o) energetic orbital set as a starting guess for the CASSCF orbital optimization and rotating the desired orbitals into the active space.  The smaller (12e,12o), (10e,10o), and (8e,8o) sets were produced by taking the (14e,14o) central-molecular orbital set and removing the most doubly-occupied (and its correlating orbital) from the active space for each size reduction. 

Notably, as shown in Figures~\ref{Figure_orbitals_1} and~\ref{Figure_orbitals_2}, the highest occupied natural orbital, CM7/E7, and the lowest unoccupied natural orbital, CM8/E8, are similar between the two orbital sets, located completely on the electrode with half occupation.
The remainder of the orbitals are quite different.
The active orbitals for the central molecular set are mainly located in the central molecular region.
The CASSCF optimization produces natural orbitals, for which  energies are not strictly defined, but upon transformation to the atomic orbital basis set and the construction of the Fock-like matrix (eq.~\ref{eq:fock}), different effective orbital energies can be found.
The HOMO-LUMO energy gap is well-known to affect conductance magnitudes, we can expect the differences in the orbital character of these two active space selections to have a noticeable impact on the conductance.

\subsection{Benzyne: Conductance vs Active Space Scaling}

\begin{figure}
\includegraphics[scale=1]{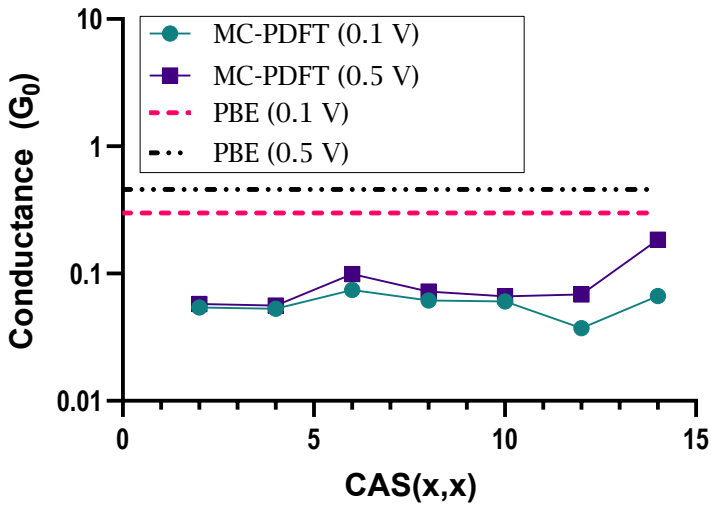}
\caption{The conductance (in quantum of conductance) versus active space size for Benzyne NEGF-MCPDFT calculations with the energetic orbital set at both 0.1 and 0.5 V. The corresponding results for the NEGF-DFT with the PBE functional are represented by the dashed lines.} 
\label{Figure_cond_benzyne}   
\end{figure}

In addition to orbital character, size constraints on the active space must typically be enforced, ideally resulting in an active space of a manageable size that can accurately determine chemical/physical properties.
Due to the large number of atoms in many organic single-molecule electronics and the computational cost of large active spaces, it would be ideal if the charge transport properties were not strongly related to the active space size and could be well-described by more modest active space sizes.
To investigate the relative importance of active 
space scaling and molecular conductance, we examined the conductance of benzyne radical using five different active space sizes and two voltages (0.1 and 0.5 volts) using the energetic orbital set.
The trends in the conductance versus active space size for both voltages are shown in Figure~\ref{Figure_cond_benzyne}.
At the lower voltage value of 0.1 V, we see relatively consistent conductance values which are below the PBE prediction by approximately one order of magnitude. The values vary from a minimum of 0.037 $G_0$ for the (12,12) active space and 0.074 $G_0$ for the (6,6) active space with a standard deviation of 0.012 $G_0$. At the higher voltage, the conductance is similarly consistent to the 0.1 V results up to (14,14) active space at which point the conductance increases noticeably to 0.19 $G_0$, half of the NEGF-DFT predicted value and noticeably higher than the other MC-PDFT active spaces. 
The 0.1 V conductance values show better overall consistency with a standard deviation of 0.012 $G_0$ vs 0.046 $G_0$ for the 0.5 V conductance. Notably, removing the (14e,14o) active space reduces the standard deviation to 0.015 $G_0$ for the 0.5V conductance suggesting that it may be a relative outlier. This increase in conductance suggests that one of poles of the (14e,14o) MC-PDFT Green's function is close to the Fermi level which we will examine in more detail in Section 4.5.

Overall, the MC-PDFT results will be relatively stable across a variety of active space sizes. This is an encouraging result and suggests a degree of active space size invariance, particularly for off-resonant, low-bias conductance. Limited variance observed between active spaces of different sizes can be seen but even the largest variance does not return the conductance value to the PBE result. For low bias voltages, it appears that only a small number of orbitals play a large role in determining the magnitude of the predicted conductance. This is in line with the behavior of non-multireference NEGF-DFT methods for which the HOMO and LUMO orbitals play a large role in determining the conductance/current at low bias voltages. Notably, the character of the CASSCF-defined MC-PDFT HOMO-LUMO orbitals is maintained across the energetic active spaces. For future NEGF-MCPDFT calculations, this scaling test can provide a systemic approach for the determination of a reasonable active space size in current/conductance calculations for molecular systems. To determine the minimal active space size needed to produce consistent transport properties, one should check the conductance values for a representative subset of voltages using a range of active space sizes and a consistent active orbital set. The stability of the conductance results vs active space for an orbital set will likely be an important metric for determining an approach active space/orbital set for a transport set.  

\subsection{Benzyne: Conductance Magnitudes}

\begin{table*}
\caption{The calculated conductance values at 0.1 V and 0.5 V for all methods and active spaces employed in this work.}
\begin{tabular}{cccccc}\hline
\label{Cond_compare}
& & \multicolumn{4}{c} {Conductance ($G/G_0$)} \\
\cline{3-6} 
Method &  Active Space & \multicolumn{2}{c}{0.1 V} & \multicolumn{2}{c}{0.5 V}  \\
\cline{3-6}
& & Ener. & Cen.-Mol. & Ener. & Cen.-Mol. \\
\hline
MC-PDFT       & (2,2)   & 0.054 &       & 0.058   &   \\
              & (4,4)   & 0.053 &       & 0.056   &    \\
              & (6,6)   & 0.074  &       & 0.099  &    \\
              & (8,8)   & 0.061  & 0.060 & 0.072  & 0.11 \\
              & (10,10) & 0.060  & 0.073 & 0.066  & 0.17\\
              & (12,12) & 0.037  & 0.099 & 0.069  & 0.24 \\ 
              & (14,14) & 0.067  & 0.034 & 0.19   & 0.44\\
PBE           &         & 0.30  &      & 0.46    & \\
\hline
\end{tabular}
\end{table*}

A simple metric for comparing the MC-PDFT transport of each active space to that of the PBE is the absolute difference in the predicted conductance values. The complete set of numerical values for benzyne conductance at 0.1~V and 0.5~V are given in Table~\ref{Cond_compare}.  
For the benzyne junction, we see notable differences between these PBE and MC-PDFT, particularly for the 0.1 V conductance values for both orbital sets. The values for both sets are between 0.034 and 0.099 $G_0$ compared to the PBE result of 0.30 $G_0$. At 0.5 V, the conductance values predicted by each orbital set separate with the energetic set predicting between 0.056-0.099 $G_0$ with the energetic active spaces from (2e,2o) to (12e,12o) and between 0.11 and 0.24 for the central-molecular active spaces. Again, the (14e,14o) active space is a notable outlier for both active spaces at 0.5 V with the 0.5 V central-molecular result nearly matching the PBE result. The results for the "central-molecular" active space overall show a greater variance in values. 
 
This further demonstrates the dependence of NEGF-MCPDFT conductance on proper active space selection and also illustrates that, despite sharing similar HOMO-LUMO characteristics, different orbital can result in  different transport characteristics, particularly at higher voltages. As this is a common characteristic of most electronic structure methods that employ an active space, it is notable that this behavior is preserved by the effective Fock matrix.  Understanding why the particular conductance results and trends in Table I were observed however requires an examination of the trends in the transmission functions. 

\subsection{Benzyne: Transmission}

\begin{figure}
\includegraphics[scale=0.8]{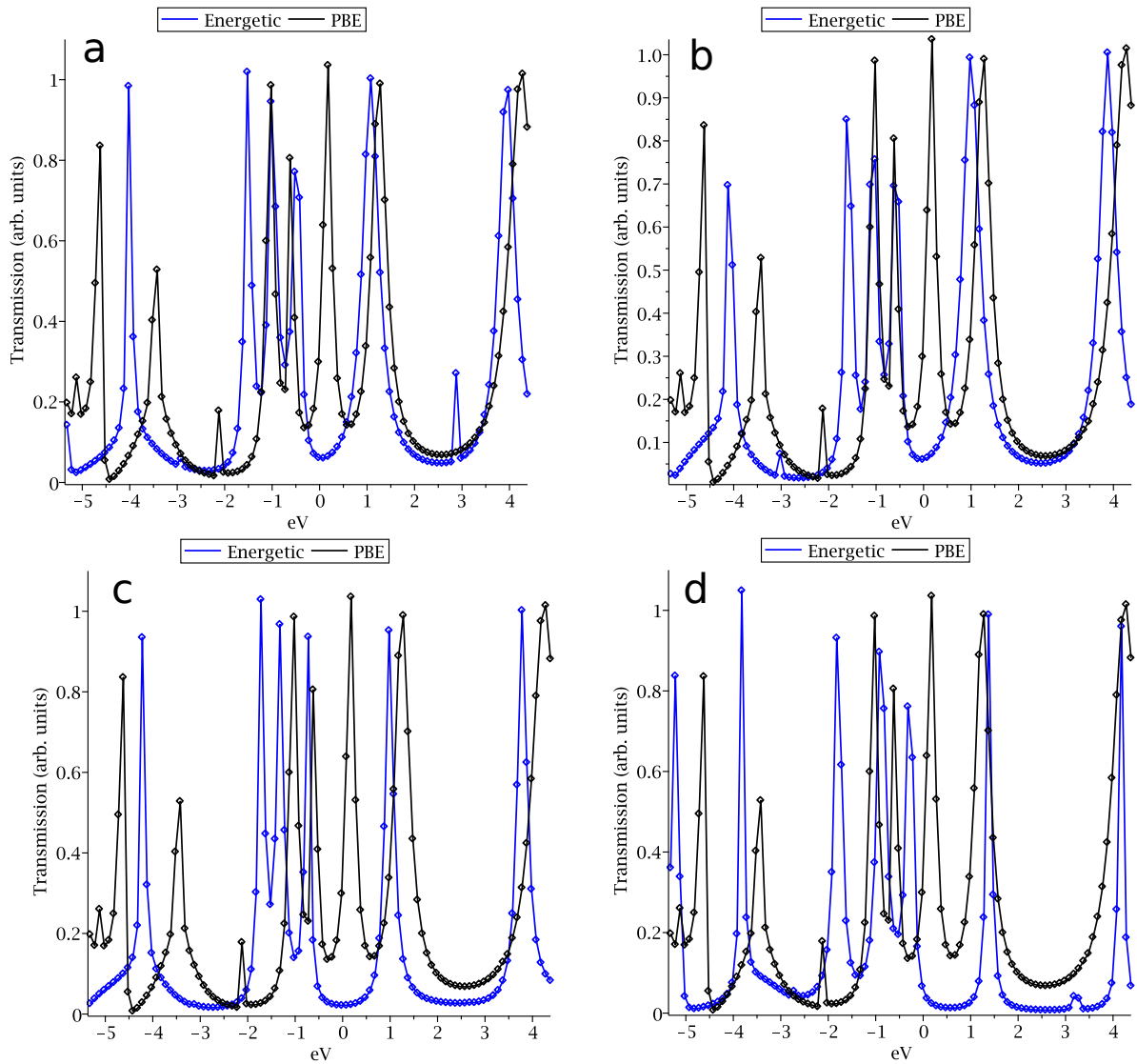}
\caption{{The transmission (arbitrary units) versus the difference from the Fermi level (eV) plots for the PBE method and MC-PDFT method with 4 energetic active space sizes: (8e,8o), (10e,10o), (12e,12o), and (14e,14o). Energetic-ordering results are in blue and PBE results are in black.}}
\label{pbe_trans}
\end{figure}

In order to better understand the cause of the conductance values seen in the previous section, we calculated transmission functions using PBE and NEGF-MCPDFT with the four largest active spaces for each orbital set.
The PBE transmission function is shown in Figure~\ref{pbe_trans}, where it is compared to the results for the energetic MC-PDFT orbital sets. All energies are relative to the approximate Fermi level which is at 0 eV in all plots. Compared to conductance values in Table~\ref{Cond_compare}, the transmission functions show clearer differences between the MC-PDFT and PBE methods. 
 The PBE functional predicts a total of four transmission peaks from -1 eV to 1.3 eV including one "double peak" around -1.02 and -0.62 eV, a large peak near the Fermi level corresponding to the PBE LUMO orbital at 0.18 eV, and the fourth at 1.28 eV above the Fermi level. The three largest peaks are roughly evenly spaced and have noticeable transmission between the peaks due to broadening effects. The MC-PDFT method using the energetic active space also predicts four peaks near the Fermi level but with lower overall transmission and shifted peak locations. The most notable difference is the large gap predicted by MC-PDFT between the 3rd and 4th peaks creating an area of low transmission near the Fermi level.  This would typically be expected to lead to lower conductance values as was consistently observed for most MC-PDFT active spaces of the energetic orbital set. This appears to primarily be a result of the large PBE transmission peak at 0.3 eV near the Fermi level being shifted to well below Fermi level for most active spaces of the energetic orbital set. This changes the transport from the resonant transport predicted by NEGF-PBE to off-resonant transport and results in the one order of magnitude reduction in the predicted conductance values seen in Table 1 for most energetic active space sizes. The notable exception is the (14e,14o) active space where the three transmission peaks below the have shifted back towards the Fermi level. The location of a transmission peak close to but not at the Fermi would result in non-linear current/voltage relationships at higher voltages and thereby a rapid increass in the predicted conductance as was observed in Table 1. Importantly, however, despite the shifted peak locations, the (14e,14o) active space predicts the same peak structure for the four peaks near the Fermi level as the other energetic active spaces with a large gap between peaks 3 and 4 rather than the even peak spacing of the PBE functional. This suggests that despite the conductance variance seen in Table 1 and Figure~\ref{Figure_cond_benzyne}, the trend is a result of the use of a fixed Fermi level rather than an issue with the MC-PDFT approach. This suggests that in future NEGF-MCPDFT studies the Fermi-level approximation should be optimized for the particular MC-PDFT active space/orbital set employed.

\begin{figure}
\includegraphics[scale=0.8]{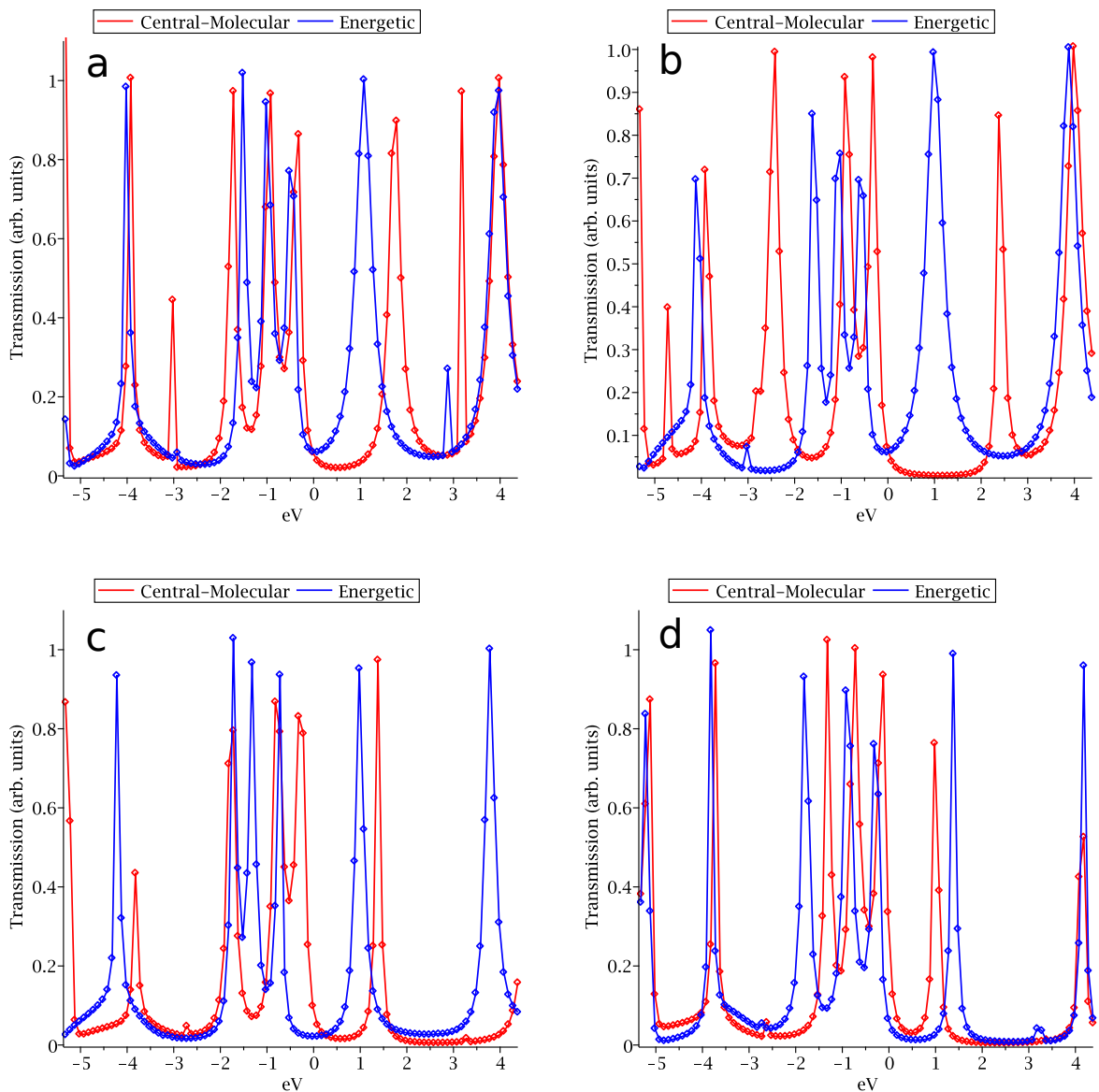}
\caption{The transmission (arbitrary units) versus the difference from the Fermi level (eV) plots for MC-PDFT method with 4 active space sizes: (8e,8o), (10e,10o), (12e,12o), and (14e,14o) in order of increasing active space size. Energetic-ordering results are in blue and central-molecular results are in red.}
\label{mcpdft_trans}
\end{figure}

Similar peak structures and energetic shifts are also evident in the central-molecular MC-PDFT orbital set as shown in Figure~
\ref{mcpdft_trans}. The central-molecular orbital set displays a similar four peak structure near the Fermi level but with more variance in the 
peak locations and relative weighting between active spaces. The 3-1 peak split structure around the Fermi level remains but the 3rd peak is 
noticeably closer to the Fermi level for the central-molecular active spaces compared to their energetic counterparts, particularly for (14e,
14o) active space which has a transmission value at the Fermi level of ~0.3, nearly identical to that of PBE at 0.30 and much larger than the 
energetic active space value of 0.07. This resulted in the large conductance increase observed for the (14e,14o) central-molecular active 
space and its similarity to the PBE conductance prediction despite the noticeable differences in the transmission functions. This again 
confirms the need to optimize the approximate Fermi level for the orbital set/active space, ideally using a self-consistent approach. From 
these results, we can see that, like other problems involving an active space approach, proper active space design will be vital to 
reproducing experimental transmission and conductance values. In addition, it is clear that the application of CASSCF orbitals is affecting the NEGF peak structure for this highly multi-reference system by altering the effective Fock matrix. This suggests that NEGF-MCPDFT is capturing significant additional electron correlation effects beyond those provided by the tPBE functional, likely including multi-reference correlation.

\section{Conclusion}

In this article, we presented a new methodology, NEGF-MCPDFT, for treating electron correlation in charge transport.
Through the use of an effective Fock matrix, MC-PDFT can replace a standard density functional within the NEGF-DFT transport formalism.
In effect, this creates a multi-reference correlation correction to the NEGF-DFT Green's functions of standard GGA functionals like PBE.
Based on our preliminary results, we found that in the cases where multi-reference correlation is not a significant factor our new methodology returns very similar results to the standard DFT functional used outside of the active space. 
When multi-reference effects are present, however, a noticeable shift in the transmission peaks and conductance values occurs as demonstrated by the benzyne calculations. 
These results have been achieved using a manageable active space size, suggesting it is probable that this MC-PDFT approach could be used for much larger molecular devices. 
The ideal active space design principles are for a molecular transport problem is, however, an open question and one that the authors intend to investigate further in the future.

\begin{acknowledgement}
A.M.S. acknowledges start-up support from Butler University.  J.T.M acknowledges the New Jersey Space Grant Consortium Training Grant A18-0090-002 which provided support through the Rowan Summer Undergraduate Research Program.  E.P.H acknowledges Rowan University which provided start-up funding for this project.

\end{acknowledgement}

\begin{suppinfo}

\end{suppinfo}

\bibliography{NEGF_MCPDFT_Paper}

\end{document}